\documentstyle[aps,prl,amssymb,epsfig]{revtex}
\begin{document}
\draft
\twocolumn[\hsize\textwidth\columnwidth\hsize\csname %
@twocolumnfalse\endcsname

\title{ Spatial Structure of Spin Polarons in the $t$-$J$ Model}
\author{A. Ram\v sak$^{1,2,3}$ and P. Horsch$^1$}
\address{$^1$ Max-Planck-Institut f\"{u}r Festk\"{o}rperforschung,
D-70569 Stuttgart, Federal Republic of Germany \\
$^2$ Faculty of Mathematics and Physics, University of Ljubljana,
SI-1000 Ljubljana, Slovenia \\
$^3$ J. Stefan Institute,  SI-1000 Ljubljana, Slovenia 
}

\date{\today}
\maketitle

\begin{abstract}
\widetext
\smallskip
The deformation of the quantum N\' eel state induced by a spin polaron
is analyzed in a slave fermion approach. Our method is based on the
selfconsistent Born approximation for Green's and the wave function for the
quasiparticle.  The results of various spin-correlation functions
relative to the position of the moving hole are discussed and shown to
agree with those available from small cluster calculations.
Antiferromagnetic correlations in the direct neighborhood of the hole
are reduced, but they remain antiferromagnetic even for $J$ as small
as $0.1t$ .  These correlation functions exhibit dipolar distortions in
the spin structure, which sensitively depend on the momentum of the
quasiparticle.  Their asymptotic decay with the distance from the hole
is governed by power laws, yet the spectral weight of the
quasiparticles does not vanish.
\end{abstract}
]

\narrowtext

\section{INTRODUCTION}

The problem of spin polarons moving in a quantum antiferromagnet has
found considerable attention, since it is important for the description of
Mott insulators at low doping\cite{rev}.
While the major part of investigations for the $t$-$J$ model
was concerned e.g. with the polaron dispersion 
and the spectral function using a variety of techniques such as
exact diagonalization\cite{hor89,bon89,hase89,Dagotto90,poi92,Leung95}, 
selfconsistent Born approximation 
(SCBA)\cite{sch88,kan89,mar91,Igarashi92}, string theory\cite{tru88,ede90} 
and other methods\cite{Prelovsek90,su89,Riera97,Barentzen96}, ---
our focus here is on the spatial structure of
the spin polarization and its asymptotic behaviour. The study of the
deformation of the spin system due to spin polaron formation was
mainly performed by exact diagonalization techniques\cite{bon89,els90}. 
However there
are important questions which can only be studied by analytical
approaches, such as the asymptotic decay of the polarization of the
medium\cite{shr88,ram93}.   
The latter property is closely related to the question whether a
quasiparticle (QP) description applies.
The first
successful measurement of the single hole dispersion in the Mott
insulator Sr$_2$CuO$_2$Cl$_2$ by angular resolved photoemission\cite{Wells95} 
has revived this interest, and stimulated investigations of
the $t$-$t'$-$J$ model\cite{tprime} 
and more complex Hamiltonians\cite{complexH}.

The Green's function for a hole moving in a {\it fixed} spin
background was discussed already in the context of transition metal oxides in
the late 60th by Bulaevskii {\it et al.} \cite{bul68} and 
by Brinkman and Rice \cite{bri70}. 
In those approaches the Green's function turned out local and fully
incoherent.  The first prediction that the low-energy single particle
excitations in the 2D $t-J$ model\cite{and87}
and its anisotropic generalization ($0\leq \alpha \leq 1$),
%
\begin{eqnarray}
H_{t-J}&=& - \,t \!\sum_{<ij>\sigma} \bigl( {\tilde c}_{i,\sigma}^\dagger 
{\tilde c}_{j,\sigma} + \mbox{H.c.} \bigr) \nonumber \\
& &+ J \sum_{<ij>}
\bigl[ S_i^z S_j^z+\frac{\alpha}{2}(S_i^+ S_j^- + S_i^- S_j^+ ) \bigr],
\label{htj}
\end{eqnarray}
are propagating quasiparticles (QP) with a bandwidth of order $J$ was
made by Kane, Lee and Read \cite{kan89} and was confirmed by a number
of exact diagonalization studies\cite{hor89,bon89}.  The problem is
complicated due to the constraint on the fermion operators $\tilde
c_{i,\sigma}^\dagger=c_{i,\sigma}^\dagger (1-n_{i,-\sigma})$ and by
the fact that quantum fluctuations play a crucial role.  This model
has been widely studied particularly because it is believed to
contain much of the low-energy physics of the high-$T_c$
superconductors\cite{and87,rev}.

Nevertheless fundamental issues are still unclear such as the
spin-dynamics and the form of the Fermi surface at moderate doping,
i.e. in the regime corresponding to the underdoped high-temperature
superconductors.  But even in the case of a single hole there are
different views e.g. whether the quasiparticle spectral weight is
finite or vanishes in the thermodynamic limit. In particular Anderson
has argued that holes introduce a deformation in the spin-background
which decays as a power law and as a consequence the spectral weight
should vanish, --- leading to non-Fermi liquid behaviour\cite{and90}.
According to this argument the non-Fermi liquid behaviour is connected
with the property of a single hole.  Recently Weng {\it et
al.} \cite{weng97} argued that the quasiparticle weight $Z_{\bf k}$
should vanish as a consequence of string formation associated with
the Marshall's sign, which is a characteristic property of the undoped
Heisenberg ground state.  These arguments are based on the appearance
of an orthogonality catastrophe in the matrix element $\langle\Psi_{\bf
k}^{exact}|c_{\bf k
\sigma}|0 \rangle$, between the exact, i.e. fully relaxed, single hole
ground state and the state $c_{\bf k \sigma}|0 \rangle$, where
$|0 \rangle$ is the ground state of the Heisenberg model without holes.

The asymptotic decay of the polarization cloud cannot be analyzed by
numerical methods, such as exact diagonalization (quantum Monte Carlo
results for the 2D t-J model are still not available), since such
studies are confined to small clusters and thus can only provide
insight into the short-range deformation of the spin-background.

A particularly powerful tool in the study of the spin polaron problem
is the slave fermion approach combined with a selfconsistent Born
approximation for the calculation of the polaron Green's 
function\cite{sch88,kan89}. This approach was successful in reproducing 
the diagonalization
results for the full Green's function obtained by
diagonalization\cite{hor89}. Therefore we shall follow this route
here. Furthermore the method properly accounts for the low-energy spin
excitations, which are crucial for the long-range distortion of the
spin-background around the moving hole. 
This method has been also applied to the finite doping 
case\cite{Plakida94,Sherman94,Plakida97}.
A further important step was
the explicit construction of the quasiparticle wave function within
the SCBA by Reiter \cite{rei94}. This wave function contains
implicitely all information about the deformation of the spin system,
and can be used to calculate this perturbation in terms of correlation
functions.

Of particular interest is here the study of relative correlation
functions (RCF), i.e. relative to the position of the hole, like for
example $C_{\bf R}=\langle n_0 
\,({\bf S}_{{\bf R}_1}\!\cdot\!{\bf S}_{{\bf R}_2})\rangle$, 
which measures the nearest neighbor correlation function for a bond 
at a distance ${\bf R}=({\bf R}_1\!+\!{\bf R}_2)/2$ from the hole at 
${\bf R}=0$
(assuming here that ${\bf R}_1$ and ${\bf R}_2$ differ by a lattice
unit vector ${\bf u}$).
Such correlation functions are usually not studied because of their
complexity. However they provide detailed information about
deformation of the spin system around the moving hole, in contrast to
the averaged correlation function 
$\langle {\bf S}_{{\bf R}_1}\!\cdot\!{\bf S}_{{\bf R}_2}\rangle$,
which measures only the global change in spin correlations due to the holes.

The results for the RCF's clearly show that the nearest neighbor spin
correlations in the neighborhood of the hole are reduced, yet
they remain antiferromagnetic (even for $J$ as small as $0.1 t$).
Therefore the frequently invoked ferromagnetic polaron picture, where
the hole is assumed to move in a ferromagnetically aligned
neighborhood of spins, does not apply to the $t$-$J$ model.

The main purpose of this work is to use Reiter's wave function for the
calculation of correlation functions
and to present a quantitative picture of the shape and size of the
quasiparticle.  While a short summary of selected results was given
earlier\cite{ram93}, the present work focuses on the description of
the technique employed for the calculation of the correlation
functions.  The technique discussed here may also be useful in other
cases where the non-crossing approximation is employed, such as more
complex models including electron-phonon coupling\cite{ram92,kyu96}.  Results for
various correlation functions describing the deformation of the
spin-background around the hole will be presented for the $t$-$J$
model [$\alpha=1$ in Eq.~(1)] as well as for the simpler $t$-$J^z$
($\alpha=0$) model\cite{Horsch94} which has no spin-dynamics and has a simple classical
N\'eel ground state.  For the $t$-$J$ model the relative correlation
functions are found to be strongly dependent on the momentum of the
quasiparticle and in good agreement with known results from exact
diagonalization.

Furthermore a detailed investigation of the asymptotic decay of
various correlation functions is given. For example the perturbation 
of the nearest-neighbor spin-correlation function $C_{\bf R}$
is found to decay as $1/R^4$ with the distance from the hole.
Since the asymptotic behaviour
of these correlation functions is closely connected with the question
whether $Z_{\bf k}$ is finite or not, it is important to calculate the
deformation of the spin-system within the different existing
approaches. In the present framework it is found that all
perturbations introduced by the hole in the quantum antiferromagnet
decay at large distance as power-law with dipolar or more complex
angular dependence depending on the momentum of the quasiparticle.
Nevertheless this does not lead to vanishing quasiparticle spectral
weight, consistent with earlier numerical results based on the study
of the polaron Green's function within the SCBA\cite{mar91}.

The plan of the paper is as follows: In Sections II and III we briefly
summarise the selfconsistent Born treatment for the Green's and the wave
function of the quasiparticle, and provide the framework 
to calculate expectation values with respect to Reiter's
wave function. Section IV deals with the quasiparticle spectral
weight, the magnon distribution function and provides a discussion of
the convergency of the approach. The more complex RCF's are studied in
Section V for two generic cases, the $t$-$J$ and the $t$-$J^z$ model,
i.e.  one with and the other without spin dynamics. This section also
contains a discussion of the asymptotic behaviour of the different
correlation functions. The paper concludes with a summary in Section
VI.

\section{SLAVE FERMION APPROACH}

In a first step of the reformulation of the problem, holes are
described as spinless (slave) fermion operators, i.e. on the
$A$-sublattice a spinless fermion creation operator is defined as
$h^+_i=c_{i\uparrow}$ while the corresponding operator
$c_{i\downarrow}=h^+_i S^+_i$ is expressed as a composite operator,
and similarly for the $B$-sublattice\cite{mar91}.  The kinetic energy
then consists of terms of form $- t h^{}_i h^+_j S^{-}_j $, 
that is, each hop of the fermion is connected with a
spin-flip.  The spin dynamics is described within linear spin wave
theory (LSW) which provides a satisfactory approximation for the 2D
spin-1/2 Heisenberg antiferromagnet.

We follow here Refs.~\cite{sch88,kan89,mar91,ram90} and express spin
operators via the Holstein-Primakoff transformation, and simplify the
notation by performing a 180$^{\circ}$ rotation of the spins on the
$B$-sublattice,
%
\begin{eqnarray}
S_i^+&=&{\case 1/2} (1+{e}^{i {\bf Q} \cdot {\bf R}_i})
(2 S-a_i^\dagger a_i)^{1/2}a_i \nonumber \\
& &+{\case 1/2} (1-{e}^{i{\bf Q} \cdot {\bf R}_i})
a_i^\dagger(2 S-a_i^\dagger a_i)^{1/2}=(S_i^-)^\dagger,  \nonumber \\
S_i^z&=&{e}^{i{\bf Q} \cdot {\bf R}_i}(S-a_i^\dagger a_i).
\label{hp}
\end{eqnarray}
Here the origin ${\bf R}_0=0$ belongs to $A$-sublattice (spin up) and
${\bf Q}=(\pi/a,\pi/a)$.  The lattice constant is $a \equiv 1$.  The
spin interaction term is further diagonalized after linearizing spin
operators and performing the Bogoliubov transformation
%
\begin{equation}
\left(\matrix{\alpha_{-{\bf q}} \cr \alpha_{{\bf q}}^\dagger
\cr} 
\right) =
\left(\matrix{u_{{\bf q}} & -v_{{\bf q}} \cr -v_{{\bf q}}
& u_{{\bf q}} \cr} \right)
\left(\matrix{b_{-{\bf q}} \cr b_{{\bf q}}^\dagger \cr} \right),
\label{alfa}
\end{equation}
where $b_{{\bf q}}^\dagger=N^{-1/2} \sum_i {e}^{-i{\bf q}
\cdot 
{\bf R}_i} a_i^\dagger$ and $N$ is the number of lattice sites.  Here
we use the usual Bogoliubov coefficients $u_{{\bf q}}$, $v_{{\bf q}}$
and the spin wave dispersion is $\omega_{{\bf q}}= 2 J \sqrt{1-(\alpha
\gamma_{{\bf q}})^2}$ with $\gamma_{{\bf q}} = (\cos q_x + \cos
q_y)/2$.  After fermion operators are decoupled into slave fermions
and bosons,
%
\begin{eqnarray}
{\tilde c}_{i\downarrow}&=&{\case 1/2} (1+{e}^{i {\bf Q} \cdot {\bf R}_i})
h_i^\dagger 
+{\case 1/2} (1-{e}^{i{\bf Q} \cdot {\bf R}_i})
h_i^\dagger S_i^-,  \nonumber \\
{\tilde c}_{i\uparrow}&=&{\case 1/2} (1+{e}^{i {\bf Q} \cdot {\bf R}_i})
h_i^\dagger S_i^-
+{\case 1/2} (1-{e}^{i{\bf Q} \cdot {\bf R}_i})
h_i^\dagger, \label{chs}
\end{eqnarray}
the fermion-magnon Hamiltonian emerges\cite{sch88}
%
\begin{equation}
{H}\!=\!{1 \over \sqrt{N}}
\!\sum_{{{\bf k}}{{\bf q}}} (
M_{{\bf k}{\bf q}} 
h_{{{\bf k}}-{{\bf q}} }^\dagger  h_{{{\bf k}} } 
\alpha_{{\bf q}}^\dagger+{\text H.c.} )
\!+\!\sum_{\bf q} \omega_{\bf q} \alpha^\dagger_{\bf q} 
\alpha_{\bf q}.
\label{lsw}
\end{equation}
A constant term irrelevant for the present discussion has been dropped
here. One recognises that the kinetic energy appears now as a
fermion-magnon coupling with a coupling function given by $M_{{\bf
k}{\bf q}}=4t(u_{\bf q} \gamma_{{\bf k}-{\bf q}} +v_{\bf q}
\gamma_{\bf k})$.  This Hamiltonian is similar to the small polaron
model except that a kinetic energy term for the spinless fermions is
absent. In the case of the cuprate superconductors, where $t>J$, the
model is in the intermediate or strong coupling regime and a
selfconsistent calculation technique must therefore be chosen.

In the following we will use the hole Green's function
%
\begin{equation}
G_{\bf k}(\omega)={1 \over \omega-\Sigma_{\bf k} (\omega)}=
{Z_{\bf k} \over {\omega-\epsilon_{\bf k}}} 
+ G^{\text inc}_{\bf k}(\omega), \label{green}
\end{equation}
where the QP band energy $\epsilon_{\bf k}$ and the pole strength
$Z_{\bf k}$ are related to the fermion self-energy $\Sigma_{\bf k}
(\omega)$ as $\epsilon_{\bf k}=
\Sigma_{\bf k}(\epsilon_{\bf k})$ 
and $Z^{-1}_{\bf k}=1- {\partial \Sigma_{\bf k}(\omega) /
\partial 
\omega}
\vert_{\epsilon_{\bf k}}$ respectively.

We calculate $\Sigma_{\bf k}(\omega)$ within the self-consistent 
Born approximation (SCBA) 
%
\begin{equation}
\Sigma_{\bf k}(\omega) = {1 \over N}\! \sum_{\bf q}  
{M^2_{{\bf k}{\bf q}}}  G_{{\bf k}-{\bf q}}( 
\omega\!-\!\omega_{\bf q}). \label{scba}
\end{equation}
Such an approximation amounts to the summation of non-crossing
diagrams to all orders. The validity of this approach is
well established.  The QP dispersion and spectral weight calculated
within the SCBA\cite{mar91} agrees very well with the exact
diagonalization results for small clusters\cite{poi92}. The spectral
weight in the limit $N\rightarrow \infty$ is finite\cite{mar91}.  In
the extreme $J/t \gg 1$ limit\cite{bul89}, 
however, this method leads to $Z_{\bf
k} \to 1$, i.e. an overestimation in comparison to $0.82$ obtained for $t=0$ in
Ref.~\cite{mal92}. The success of the SCBA has roots in the vanishing
of low order vertex corrections as pointed out by
several authors for systems where the hole is coupled to an AFM spin
background\cite{ram90,mar91,liu92}.

\section{QUASIPARTICLE WAVE FUNCTION}

Given the Green's function in selfconsistent Born approximation 
it would be interesting to have the wave function of the quasiparticle
which corresponds to the pole in Eq.~(\ref{green})
at energy $\epsilon_{\bf k} =
\Sigma_{\bf k} (\epsilon_{\bf k})$.  The knowledge of this wave
function will allow us to calculate in principle all equal-time
correlation functions which define the perturbation of the
AFM-background around the hole.
We follow here closely Reiter's \cite{rei94} original approach and
prove in addition that the quasiparticle weight derived from the wave
function is consistent with the well known expression obtained from
the Green's function.

The quasiparticle wave function is defined as the eigenstate of $H$
%
\begin{equation}
H |\Psi_{\bf k}\rangle = \epsilon_{\bf k} |\Psi_{\bf k}\rangle, \label{e8}
\end{equation} 
which gives rise to the quasiparticle peak in the spectral
representation for the Green's function 
%
\begin{equation}
G_{\bf k} (\omega) = \sum_m \frac{\mid \langle \Psi_{{\bf k}m} 
\mid h_{\bf k}^+ \mid 0 \rangle \mid^2}
{\omega - \epsilon_{{\bf k}m}}. \label{green1}
\end{equation}
Here $|0\rangle$ represents the vacuum state with respect to fermion
and magnon operators and $|\Psi_{{\bf k}m} \rangle$ is an eigenstate of
Hamiltonian Eq.~(\ref{lsw}) with eigenenergy $ \epsilon_{{\bf k}m}$.
The spectral weight of the quasiparticle state 
$|\Psi_{{\bf k}} \rangle$
%
\begin{equation}
Z_{\bf k} = \mid \langle \Psi_{\bf k} \mid h_{\bf k}^+ \mid 0
\rangle \mid^2
\end{equation}
can be quite small, however it should not scale to zero in the
thermodynamic limit, whereas the matrix elements contributing to the
incoherent part are of $O(1/N$) or smaller.

Given the Hamiltonian Eq.~(\ref{lsw}) we expect the quasiparticle wave 
function $\mid \Psi_{\bf k}\rangle$  to have the form
\begin{eqnarray}
\mid \Psi_{\bf k} \rangle& = & a^0 ({\bf k}) h_{\bf k}^+ |0\rangle  
+  \frac{1}{\sqrt{N}}
\sum_{{\bf q}_1} a^1 ({\bf k},{\bf q}_1) h_{{\bf k}-{\bf q}_1}^+ 
\alpha_{{\bf q}_1}^{+}|0\rangle \nonumber\\
& + & \frac{1}{N} \sum_{{\bf q}_1{\bf q}_2} a^2 ({\bf k},{\bf
q}_1,{\bf q}_2) h_{{\bf k}-{\bf q}_1 - {\bf q}_2}
\alpha_{{\bf q}_2}^+ \alpha_{{\bf q}_1}^+ |0\rangle\nonumber\\ 
& + & ... ,\label{eq11}
\end{eqnarray}
where the coefficients $a^n ({\bf k},{\bf q}_1, ...,{\bf q}_n)$ are to be
determined. 

>From the Schr\"odinger equation we obtain the following
system of equations for the expansion coefficients:
\begin{equation}
\omega  a^0 ({\bf k}) - \frac{1}{N} \sum_{{\bf q}_1} a^1 ({\bf k},{\bf
q}_1) M_{{\bf kq}_1} =  0  \label{eq12}
\end{equation}
and
\begin{eqnarray}
\Bigl( \omega -\omega_{{\bf q}_1} \Bigr) a^1 ({\bf k},{\bf q}_1) 
- a^0 ({\bf k}) M_{{\bf k},{\bf q}_1}\;\;\;\; & & \nonumber\\  
- \frac{1}{N}  \sum_{{\bf q}_2} a^2 ({\bf k},{\bf q}_1,{\bf q}_2) 
M_{{\bf k}-{\bf q}_1,{\bf q}_2} & = & 0. \label{eq13}
\end{eqnarray}
\noindent 
To obtain these equations which correspond to the noncrossing
approximation for the Green's function one has to adopt the following
contraction rule: When one magnon is annihilated in the
$n$-magnon component of the wave function, Eq.~(\ref{eq11}), only the
contribution is considered where the last magnon in the sequence,
i.e. $\alpha_{{\bf q}_n}^+$, is annihilated.  This is reminescent of
the retraceable path approximation in momentum space. The general
equation for $n > 0$ reads:
\begin{eqnarray}
\Bigl( \omega - \omega_{{\bf q}_1} &...& -  \omega_{{\bf q}_n} 
\Bigr) a^n ({\bf k}, ..., {\bf q}_n) 
- a^{n-1} ({\bf k},..., {\bf q}_n) M_{{\bf k}_{n-1},{\bf q}_n} \nonumber\\
& - & \frac{1}{N} \sum_{{\bf q}_{n+1}} a^{n+1} ({\bf k}, ..., {\bf
q}_{n+1}) 
M_{{{\bf k}_n},{\bf q}_{n+1}} = 0 , \label{eq14}
\end{eqnarray}
where ${\bf k}_n = {\bf k}-{\bf q}_1 -...-{\bf q}_n$.

As first shown by Reiter \cite{rei94} this sequence of equations 
(\ref{eq12})-(\ref{eq14}) has the general solution

\begin{equation}
a^{n+1} ({\bf k},...,{\bf q}_{n+1}) = a^n ({\bf k}, ..., {\bf q}_n) 
g_{{\bf k}_n,{\bf q}_{n+1}} \ , \label{eq15}
\end{equation}
where we introduced the abbreviation
\begin{equation}
g_{{\bf k}_n,{\bf q}_{n+1}} = M_{{\bf k}_n,{\bf q}_{n+1}} G_{{\bf k}_{n+1}}
 (\omega-\omega_{{\bf q}_1} - ... - \omega_{{\bf q}_{n+1}}) \ . 
\label{eq16}
\end{equation}

Substituting Eq.~(\ref{eq15}) into the last term on the $l.h.s.$ of 
Eq.~(\ref{eq14}), we
recognise that this term is identical to the expression Eq.~(\ref{scba})
for the selfenergy
$\Sigma_{\bf k}(\omega)$ times $a^n$. This yields for Eq.~(\ref{eq14})

\begin{eqnarray}
\Bigl( \omega - ... - \omega_{{\bf q}_n} & - & \Sigma_{{\bf k}_n} 
(\omega - ... - \omega_{{\bf q}_n}) \Bigr) a^n ({\bf k}, ...,{\bf
q}_n) \nonumber\\ 
& - & a^{n-1}  ({\bf k}, ..., {\bf q}_{n-1}) M_{{\bf k}_{n-1},{\bf q}_n} = 0.
\end{eqnarray}
Since the prefactor of $a^n$ is the inverse of the Green's function
$G_{{\bf k}_{n}}(\omega -\omega_{{\bf q}_1} - ... - \omega_{{\bf q}_{n}})$ 
this equation is identical to Eq.~(\ref{eq15}) with $n$ replaced by $n-1$.
It only remains to be shown that also Eq.~(\ref{eq12}) is solved. 
Equation (\ref{eq12}) becomes

\begin{equation}
a^0({\bf k}) \bigl( \omega - \Sigma_{\bf k} (\omega) \bigr) = 0 \ ,
\end{equation}
which has a nontrivial solution $a^0({\bf k}) \neq 0$ at the QP-energy
$\omega = \epsilon_{\bf k}$.  The knowledge of the Green's function
Eq.~(\ref{green}) is sufficient to calculate from Eq.~(\ref{eq15})
iteratively the coefficients $a^n ({\bf k},{\bf q}_1, ... {\bf q}_n)$. 

The coefficient $a^0({\bf k})$ which determines the
QP-weight $Z_{\bf k} = [a^0({\bf k})]^2$ follows from the 
normalisation of the wave function
$\langle \Psi_{\bf k} \mid \Psi_{\bf k} \rangle  =  
\sum_{n=0}^\infty \mid a^n ({\bf k}, ...  ,{\bf q}_n)
\mid ^2 = 1 $,

\begin{eqnarray}
 \langle \Psi_{\bf k} \mid \Psi_{\bf k} \rangle 
& = & [a^0({\bf k})] \biggl\{ 1 + \frac{1}{N} \sum_{{\bf q}_1} 
g^2_{{\bf k},{\bf q}_1} +  \label{eq19} \\
& & +  \frac{1}{N^2} \sum_{{\bf q}_1{\bf q}_2} 
g^2_{{\bf k},{\bf q}_1} g^2_{{\bf k}-{\bf q}_1,{\bf q}_2} 
 +  \biggl. \ ... \biggr\}.\nonumber
\end{eqnarray}
When one calculates the derivative $\partial \Sigma_{\bf k}
(\omega)/\partial \omega$ from Eq.~(\ref{scba}) and compares the result
with Eq.~(\ref{eq19}) it is easy to see that\cite{ram93}
\begin{equation}
\langle \Psi_{\bf k} \mid \Psi_{\bf k}\rangle = 
[a^0({\bf k})]^2 
\left( 1 - \frac{\partial 
\Sigma_k (\omega)}{\partial \omega} \right) _{\omega =\epsilon_k} \ .
\end{equation}
%
%
%
As $|\Psi_{\bf k}\rangle$ is normalized to 1, $[a^0({\bf k})]^2$ is
indeed identical to the QP-spectral weight as calculated directly from
$G$.  This latter step is important, since it accomplishes the prove
of the internal consistency of $G$ and $\Psi$, i.e. where both are
calculated within selfconsistent Born approximation.  It should be
emphasised that the above derivation does not rely on the assumption
that the coupling term in the Hamiltonian is small.

Because of the presence of AFM long-range order the quasiparticles
move on one sublattice, while visiting the other sublattice only
virtually. In view of the 'degeneracy' $\epsilon_{{\bf k}+{\bf
Q}}=\epsilon_{\bf k}$ and $G_{{\bf k}+{\bf Q}}=G_{\bf k}$ we define
new Bloch-operators
\begin{equation}
h_{{\bf k} \tau}^\dagger=2^{-1/2}
(h_{\bf k}^\dagger+\tau h_{{\bf k}+{\bf Q}}^\dagger) \label{htau}
\end{equation}
which create holes on the $\uparrow$ ($\downarrow$) sublattice for
$\tau=1(-1)$, respectively. The momenta ${\bf k}$ are now restricted
to the reduced (AFM) Brillouin zone.  The corresponding wave functions
including magnon operators up to order $n$ are
%
\begin{equation}
|\Psi_{{\bf k}\tau}^{(n)}\rangle = \frac{1}{\sqrt{2}}
\Bigl( |\Psi_{\bf k}^{(n)}\rangle  +
\tau |\Psi_{{\bf k}+{\bf Q}}^{(n)}\rangle \Bigr) \label{psitau}
\end{equation}
with $|\Psi_{{\bf k}}^{(n)}\rangle$ following from Eq.~(\ref{eq11})
%
\begin{eqnarray}
|\Psi_{{\bf k}}^{(n)}\rangle&=&Z_{\bf k}^{1/2} \Bigl[
\;h_{{\bf k}}^\dagger 
+N^{-1/2}\sum_{{\bf q}_1} g_{{\bf k},{\bf q}_1} 
h_{{\bf k}_1}^\dagger \alpha_{{\bf q}_1}^\dagger+\;..... \nonumber\\
& + & N^{-n/2}\!\!\!\sum_{{\bf q}_1,.....,{\bf q}_n} \!\!
g_{{\bf k},{\bf q}_1} g_{{\bf k}-{\bf q}_1,{\bf
q}_2}\; 
... \label{psi} \\
& \times & g_{{\bf k}-{\bf q}_1-...-{\bf q}_{n-1},
{\bf q}_n} h_{{\bf k}-{\bf q}_1-...-{\bf q}_n}^\dagger
\alpha_{{\bf q}_1}^\dagger...\,\alpha_{{\bf q}_n}^\dagger \Bigr]
|0\rangle. \nonumber 
\end{eqnarray}
Here $g_{{\bf k},{\bf q}}=M_{{\bf k}{\bf q}} G_{{\bf k}-{\bf q}}
(\epsilon_{\bf k}-\omega_{\bf q})$ as defined in Eq.~(\ref{eq16}).  We
note that the Green's function $G$ in $g_{{\bf k},{\bf q}}$ is always
evaluated below the lowest pole and therefore real. For example in $
G_{{\bf k}-{\bf q}}(\epsilon_{\bf k}-\omega_{\bf q})$ the energy
$\epsilon_{\bf k-q} > \epsilon_{\bf k}-\omega_{\bf q}$, hence $g_{{\bf
k},{\bf q}}$ is real. This is actually a subtle consequence of the
selfconsistent evaluation of the Green's function which leads to a
smaller energy variation of the QP energy compared to the spin wave
dispersion\cite{mar91} .  We stress that this holds also true in the
strong coupling case where also the $\epsilon_{\bf k}$ variation is of
order $J$, i.e. comparable with spin wave energies.  The choice of
sublattice wave functions Eq.~(\ref{psitau}) is convenient since they
are eigenstates of $S^z_{tot}$ with eigenvalues $\pm 1/2$.

The diagrammatic structure of the wavefunction $|\Psi_{{\bf
k}\uparrow}^{(n)}\rangle$ is shown in Fig.~1(a).  The translation
rules are straightforward: (1) Open ends on the right correspond to
operators $ h_{{\bf k}-{\bf q}_1-...-{\bf q}_n}^\dagger$ and
$\alpha_{{\bf q}_1}^\dagger...\,\alpha_{{\bf q}_n}^\dagger$, (2) thin
lines are associated with $\sqrt{Z_{\bf k}}$, (3) a vertex (dot)
connected with a double line corresponds to $g_{\bf k,q}$, and (4)
there is a momentum sum for each magnon line.  It is obvious that the
wavefunction does not correspond to a strict order $n$ expansion with
respect to the fermion-magnon coupling, since the Green's function,
Fig.~1(b), is already evaluated selfconsistently with respect to this
interaction.

In the next Section we will investigate the relative importance of the
different terms in the wave function Eq.~(\ref{psi}) and address the
question under which conditions this series can be truncated.

\section{MAGNON DISTRIBUTION FUNCTION}

The first question one may ask is: ``How many magnons are involved in
the formation of the polaron?'' As the coupling between hole and
spin-excitations is the kinetic energy of the $t$-$J$ model, small
values of $J/t$ correspond to strong coupling (small spin stiffness),
where many magnons are excited by the hole motion.  In order to
estimate the number $n$ of magnon terms needed in the wave function we
have calculated the norm ${\cal N}_{\bf k}$
%
\begin{equation}
{\cal N}_{\bf k}
=\langle\Psi_{{\bf k} \tau}^{(n)} 
|\Psi_{{\bf k} \tau}^{(n)}\rangle=
\sum_{m=0}^n A_{\bf k}^{(m)}. \label{nak}
\end{equation}
The distribution function $ A_{\bf k}^{(m)}$ defines the probability
for the $n$-magnon contribution in the wave function. A similar study
on a small cluster was presented in Ref.~\cite{dag91}.  

In Fig.~2 the norm is presented diagrammatically consistent with
Eq.~(\ref{eq19}).  Each term $A_{\bf k}^{(m)}$ corresponds to a single
non-crossing diagram with $n$-magnons. Vertices denoted with dots
correspond to the fermion-magnon coupling matrix elements $M_{\bf k
q}$ and the double line to the {\it square} of Green's function
Eq.~(\ref{green}) calculated within SCBA.  The analytical expression for
$A_{\bf k}^{(m)} $ is independent of $\tau$ and given by
%
\begin{equation}
A_{\bf k}^{(m)}={Z_{\bf k} \over N^{m}} \sum_{{\bf
q}_1,...,
{\bf q}_m}
\!\!g_{{\bf k},{\bf q}_1}^2 g_{{\bf k}-{\bf
q}_1,{\bf 
q}_2}^2 \;...\;
g_{{\bf k}-{\bf q}_1-...,{\bf q}_m}^2 \label{aterms}
\end{equation}
for $m>0$, while
$A_{\bf k}^{(0)}=Z_{\bf k}$. From  Eq.~(\ref{eq19}) we know
that ${\cal N}_{\bf k} \to 1$ in the limit $n\to \infty$. This
normalisation condition will serve as a check of our numerical procedure.

It is instructive to study the distribution function $A_{\bf k}^{(m)}$
first in the case of the the $t$-$J^z$ model ($\alpha=0$).  In this
limit of model Eq.~(\ref{htj}) the analysis becomes simple because there
is no intrinsic spin-dynamics.  The SCBA equations for the self-energy
are independent of ${\bf k}$ and reduce to one equation $\Sigma_{\bf
k}(\omega)=4t^2[\omega-2J^z-\Sigma_{\bf k} (\omega-2J^z)]^{-1}$
\cite{kan89}.
Equation (\ref{aterms}) can then be expressed in a recurrence form
%
\begin{equation}
A_{\bf k}^{(m+1)}=A_{\bf k}^{(m)}[2t G_{\bf k}(\epsilon_{\bf 
k}-2mJ^z)]^2.
\end{equation}
The norm ${\cal N}_{\bf k}$ is shown in Fig.~3 as a function of the
number of magnon terms $n$ for various $J^z/t$.  A crossover between
the weak and the strong coupling regime occurs at $J^z/t \sim
0.3$. For smaller $J^z/t$ the number of magnon terms needed to fulfil
the sum rule ${\cal N}_{\bf k}=1$ increases rapidly. In Fig.~4 the
distribution of magnons $A_{\bf k}^{(m)}$ is displayed for the strong
coupling case, $J^z/t
\ll 1$ .  In this regime $A_{\bf k}^{(m)}$ has a
maximum at a finite value $n$, which increases with the coupling
constant $t/J^z$.

The average number of magnons forming the spin polaron may be defined as
%
\begin{equation}
\langle n \rangle=
\langle\Psi_{{\bf k} \tau}^{(n)}|
\sum_{\bf q} \alpha_{\bf q}^\dagger \alpha_{\bf q}
|\Psi_{{\bf k} \tau}^{(n)}\rangle=
\sum_m m A_{\bf k}^{(m)}. \label{nmag}
\end{equation}
In the Ising limit $\langle n \rangle $ is identical to the average
number of spin deviations (local magnons) $\langle \sum_i S_i^+ S_i^-
\rangle= \langle \sum_i a_i^\dagger a_i \rangle$.  It is evident that
the latter expression is proportional to the average string length
$l_{av}$ \cite{shr88} of overturned spins in the N\' eel state created
by the hole motion. As the string potential is an approximately linear
function of the string length this implies $\langle n \rangle \propto
l_{av} \propto (t/J^z)^{1/3}$.  This estimate is reasonable for long
strings, i.e. small $J^z/t$.  In Fig.~5 we present $\langle n \rangle$
as a function of $J^z/t$ calculated with up to 40 magnon terms in the
wave function. For large $J^z/t \gg 1$ only the leading term $m=1$ in
Eq.~(\ref{nmag}) is relevant, therefore the asymptotic result is
$\langle n \rangle=(t/J^z)^2$. For $J^z/t \ll 1$ we find excellent
agreement with the result $\langle n \rangle=1.4 (t/J^z)^{1/3}$
obtained by Mattis and Chen~\cite{mat91}.

From these results for the $t$-$J^z$ model it is clear for $J^z/t \geq
0.4$ the wave function can be truncated at $n=3$ or even at $n=2$.  We
note that the same holds true for the $t$-$J$ model\cite{ram93}.

In Figures 6(a) and 6(b) the numerical results for the norm ${\cal
N}_{\bf k}$ of the $t$-$J$ model are shown for ${\bf k}=0$ and ${\bf k}=
(\frac{\pi}{2},\frac{\pi}{2})$, respectively, both calculated with up
to $n=3$ magnons kept in the wavefunction. For $J=0.4$ 3-magnon
contributions are necessary to fulfil the norm. The quasiparticle
spectral weight $Z_{\bf k}$, which corresponds to the $n=0$ term
displayed in Figs. 6(a,b), is always finite except in the limit of
vanishing spin stiffness $J=0$.  Thus our wavefunction does not lead
to an orthogonality catastrophe. This result will be further
complemented in the next Section by a detailed study of the asymptotic
decay of the spin-polaron correlation functions.

The question whether the QP spectral weight $Z_{\bf k}$ for the
$t$-$J$ model is finite or not is still not completely
settled. Numerical results obtained on small clusters are in a good
agreement with the results obtained from the SCBA 
Equations~(\ref{scba})\cite{mar91}.  In the SCBA formalism 
Eq.~(\ref{lsw}), $Z_{\bf k}$ is
finite\cite{kan89,ram90,mar91} because the hole-magnon coupling
matrix element for $q \to 0$ is not singular and therefore the hole is
weakly coupled to low energy spin waves.  In Ref.~\cite{weng97} it was
argued that $Z_{\bf k}$ should vanish nevertheless because of string
like phases associated with the hole motion (due to hidden Marshall
signs). We stress that the Marhall sign convention is implicitely
included in our present formulation. In fact the 
vacuum state $|0\rangle$ (originating via unitary transformation) is
equivalent to the quantum N\' eel groundstate of the $T=0$ Heisenberg
model, and thus by construction obeys the Marshall sign rule in the original
basis, i.e. before the 180$^\circ$ rotation of the $B$-sublattice. After the
transformation, Eq.~(\ref{hp}), there are no additional phases in the 
transformed Hamiltonian due to the Marshall sign.

Not considered in the present treatment, is the effect of the 4 broken
bonds meeting at the site of the hole. This leads to an additional
relaxation of the spin correlations and hence to a reduction of the
quasiparticle weight. This effect is expected to be strongest in the
limit $t=0$. The exact result for the spectral weight in this case is
$Z=0.82$ and was derived by Mal'shukov and Mahan \cite{mal92} (as
compared to 1 in the present treatment). The energy change due to the
broken bonds must also be included in the Born approximation if one
wants to compare the quasiparticle energies with those from exact
diagonalization, as discussed by Mart\'{\i}nez and Horsch \cite{mar91}.

\section{SPIN-POLARON CORRELATION FUNCTIONS}

The spatial structure of the spin-polaron can be described with
various correlation functions measuring the perturbation of the spin
system relative to the position of the moving hole.  As we shall see,
these correlation functions are strikingly different in the $t$-$J$
and the $t$-$J^z$ model, --- a consequence of the absence of spin
dynamics in the latter model.  In the $t$-$J$ model perturbations
created by the hole are carried away by spin waves thereby generating
a power law perturbation pattern with an interesting angular
dependence, whereas in the absence of spin dynamics the perturbations
are characterized by an isotropic gaussian decay.

Such relative correlation functions can be evaluated using the
quasiparticle wave function. One of the simplest correlation functions
is the distribution of magnons around the hole
%
\begin{equation}
N_{\bf R}=
\langle\Psi_{{\bf k} \tau}^{(n)}|
\sum_i n_i a_{{\bf R}_i+{\bf R}}^\dagger a_{{\bf R}_i+{\bf R}}
|\Psi_{{\bf k} \tau}^{(n)}\rangle\equiv 
\langle n_0 (a_{\bf R}^\dagger a_{\bf R})\rangle. \label{nr}
\end{equation}
Here $n_i=h_i^\dagger h_i$ is density operator for holes at site $i$
with position ${\bf R}_i$.  $N_{\bf R}$ also corresponds to the
distribution of spin deviations, $\langle n_0 (S_{\bf R}^+ S_{\bf
R}^-)\rangle$. Therefore it provides a suitable measure of the polaron
size. This correlation function is also proportional to the
distributions $\langle n_0 (S_{\bf R}^x)^2\rangle=\langle n_0 (S_{\bf
R}^y)^2\rangle$.

Correlation functions such as $N_{\bf R}$ are evaluated using
similar diagrams as in the calculation of the norm ${\cal N}_{\bf k}$.   
One has to evaluate the expectation values
%
\begin{equation}
\langle\Psi_{{\bf k} \tau}^{(n)}|
\sum_i n_i \hat {\cal O}_{{\bf R}_i+{\bf R}} 
|\Psi_{{\bf k} \tau}^{(n)}\rangle\equiv 
\langle n_0 \hat {\cal O}_{{\bf R}}
\rangle. \label{hato}
\end{equation}
Here the summation $\sum_i$ corresponds to all lattice sites and
the density operator for the hole
%
\begin{equation}
n_i={1 \over N} \sum_{{\bf k}_1 {\bf k}_2}   
e^{i ({\bf k}_2-{\bf k}_1)\cdot {\bf R}_i} 
h^\dagger_{{\bf k}_1} h_{{\bf k}_2}
\end{equation}
has to be expressed in terms of operators $h_{{\bf k} \tau}$,
Eq.~(\ref{htau}). The operator $\hat {\cal O}_{\bf R}$ is decomposed
into magnon variables as
%
\[
\hat {\cal O}_{\bf R}={1 \over N} \sum_{{\bf q}_1 {\bf q}_2}\Bigl[
f_{{\bf q}_1 {\bf q}_2} ({\bf R})
\alpha_{{\bf q}_1}^\dagger \alpha_{{\bf q}_2} 
+g_{{\bf q}_1 {\bf q}_2} ({\bf R})
\alpha_{{\bf q}_1} \alpha_{{\bf q}_2}+
{\text H.c.} \Bigr].
\]
%
The diagrammatic structure of the contributions for a general
correlation function of this type is presented in Fig.~7.  The first
class of diagrams is symmetric and derives from the vertex function
$f_{{\bf q}_1 {\bf q}_2} ({\bf R})$.  These diagrams, denoted by
$(B_n)$, arise as diagonal contributions from the $n$-magnon component
of the wave function.

The construction rule for these diagrams is the
following: If the vertex $f$ (circle) as well as the connected two
magnon lines (together with their vertices and associated double
lines) are removed from the diagram, one must arrive at a diagram
contained in the expression for the norm (Fig.~2). Otherwise the
diagram is not consistent with the selfconsistent Born approximation
and should be dropped.

The second class of diagrams $(C_{nm})$ is asymmetric and corresponds to
the vertex function $g_{{\bf q}_1 {\bf q}_2} ({\bf R})$ which connects
$n$ magnon contributions with $m=n\pm 2$ magnon terms in the wave
function. Again only such diagrams must be taken into account which
are consistent with the construction rule formulated before.

The vertex functions $f$ and $g$ are expressed in terms of Bogoliubov
coefficients and thus strongly momentum dependent. For the case of the
correlation function $N_{\bf R}$ we have
%
\begin{eqnarray}
f_{{\bf q}_1 {\bf q}_2} ({\bf R})&=&
{\case 1/2} (u_{{\bf q}_1} u_{{\bf q}_2}+v_{{\bf q}_1} v_{{\bf q}_2}  )
e^{i({{\bf q}_1-{\bf q}_2})\cdot {\bf R}} \nonumber\\
g_{{\bf q}_1 {\bf q}_2} ({\bf R})&=&
{\case 1/2} (u_{{\bf q}_1} v_{{\bf q}_2}+v_{{\bf q}_1} u_{{\bf q}_2}  )
e^{i({{\bf q}_1+{\bf q}_2})\cdot {\bf R}}. \label{fg}
\end{eqnarray}
In order to illustrate a typical calculation of matrix elements needed
in the correlation functions, we present here the second order
contributions $B_2$ in Fig.~7,
\begin{eqnarray}
B_2&=& N^{-3} \!\!\sum_{{\bf q}_1 {\bf q}_2 {\bf q}_3} 
\!f_{{\bf q}_1 {\bf q}_2} ({\bf R}) (
g_{{\bf k},{\bf q}_3}g_{{\bf k}-{\bf q}_3,{\bf q}_1}
g_{{\bf k}-{\bf q}_3,{\bf q}_2}g_{{\bf k},{\bf q}_3}
+ \nonumber \\
& + &g_{{\bf k},{\bf q}_1}g_{{\bf k}-{\bf q}_1,{\bf q}_3}
g_{{\bf k}-{\bf q}_2,{\bf q}_3}g_{{\bf k},{\bf q}_2} ),  \label{b2}
\end{eqnarray} 
where the first and the second term correspond to noncrossing and
crossing term, respectively.

\subsection{Ising limit ($\alpha=0$)}

In general correlation functions and the corresponding 
matrix elements have to be evaluated numerically, which
is easy for not too large $n$ up to $\sim5$. The $t$-$J^z$ model is
an exception, since the Bogoliubov factors simplify to
$u_{\bf q}=1$ and $v_{\bf
q}=0$. Thus Re $f_{{\bf q}_1 {\bf q}_2} ({\bf R})=
\cos {({{\bf q}_1-{\bf q}_2})\cdot {\bf R}}$ and
$g_{{\bf q}_1 {\bf q}_2} ({\bf R})=0$, respectively. 
As the Green's function is dispersionless
it is possible to express the matrix elements analytically
and perform the summation of diagrams $(B_n$) to any order. 
Furthermore diagrams $(C_{nm})$ are zero. It is
instructive to express the wave function in real space.
Each $n$-magnon term can then be visualized as a string
of $n$ steps with starting point at the origin. 
From such a study one can get insight into the noncrossing
structure of the wave function and correlation functions.

The SCBA is similar to the retraceable path approximation, yet with
the important difference that in SCBA the hole can also hop backwards
on its path. At the level of the Green's function the differences were
discussed in Ref.~\cite{mar91}.  The result for the magnon
distribution function, Eq.~(\ref{nr}) can therefore be expressed as
%
\begin{equation}
N_{\bf R}=\sum_{m=1}^n p_m({\bf R}) P_m, \label{npp}
\end{equation}
where $P_m=\sum_{j=m}^n A_{\bf k}^{(j)}$ can be interpreted as a
probability to have at least $m$ local magnons excited. The
coefficients $p_m({\bf R})$ represent the probability that a string of
$m$ excited local magnons ends at a given lattice position ${\bf R}$.
This distribution 
can be determined by counting all possible paths of $m$ steps, 
where in each step all $z$ neighbors can be reached, 
%
\begin{equation}
p_m({\bf R})=4^{-m}{ m\choose m_+}{ m\choose m_-}. \label{pmr}
\end{equation}
Here $m_\pm=(m-\big| |R_x| \pm |R_y|\big|)/2$
must be a non-negative integer, otherwise $p_m({\bf R})=0$.
This result is free of boundary conditions. 

The correlation function $N_{\bf R}$ can be used to determine the
spatial size of the polaron in the Ising limit. We define the size of
the polaron quantitatively by the radius $R_p$ (element of Bravais
lattice), which encloses a given fraction $p$ of the total number of
spin deviations, $p=\langle n \rangle^{-1}\sum_{R\leq R_p}
N_{\bf{R}}$. In Fig.~8 the polaron radius $R_p$ vs. $J/t$ is shown for
three different values of $p=0.75$, $0.9$, and $0.99$. In the
physically interesting regime, $J^z/t \sim 0.3$, the polaron is
contained within the radius $R<2$.  The scaling $R_p\propto \langle n
\rangle^{1/2}
\propto (t/J^z)^{1/6}$ expected for the polaron\cite{shr88,ram93}
is well established. We have also calculated the average radius,
$\langle R \rangle=\langle n\rangle ^{-1}\sum_{\bf R} |{\bf R}|
N_{\bf{R}}$, and the root-mean square radius, $R_{RMS}= (\langle n
\rangle ^{-1}\sum_{\bf R} |{\bf R}|^2 N_{\bf{R}})^{1/2}$.  In Fig.~8
$R_{RMS}$ and $\langle R\rangle$ are presented with solid and dashed
lines, respectively. The RMS radius can be well fitted with $R= 1.06
(t/J)^{0.157}$ for $J^z/t<1$.

In the Ising limit the total spin is not conserved. However, the
$z$-component of spin is a conserved quantity. A state with one static
hole ($t=0$) at the site $i_0$ has by definition the $z$-component of
total spin $S_{tot}^z=\sum_{i\ne i_0}S_i^z=-\tau/2$ ($\tau=\pm 1$),
i.e. the spin of one site of the sublattice not corresponding to
$i_0$. If the hole becomes mobile ($t\ne 0$), some spins around the
hole deviate from the N\' eel order. The region where the spin order is
disturbed corresponds to the spin polaron defined above. The
correlation function describing the spatial distribution of spin
around the hole is thus
%
\begin{eqnarray}
S_{\bf R}&=&\langle n_0 S_{\bf R}^z \rangle= \nonumber\\
&=&\tau e^{i {\bf Q}\cdot{\bf R}} \sum_i 
e^{i {\bf Q}\cdot{\bf R}_i} \Bigl[
{\case 1/2} \langle n_i \rangle-\langle n_i a_{{\bf R}_i+{\bf R}}^\dagger
a_{{\bf R}_i+{\bf R}}\rangle \Bigr], \label{sr}
\end{eqnarray}
where we have expressed spin operators in terms of magnons according
to Eq.~(\ref{hp}). The conservation of spin corresponds to the
sum rule $\sum_{R\ne 0}S_{\bf R}=-\tau/2$.  The local spin operator is
within the LSW approximation related to the number of bosons, $S_i^z =
{\case 1/2}-a_i^\dagger a_i$. However, $S_{\bf R}$ is due to the
factor $\exp({i {\bf Q}\cdot{\bf R}_i})$ non-trivially related to
$N_{\bf R}$ and has to be calculated independently.  After carrying
out the steps similar as in the evaluation of $N_{\bf R}$ one obtains
%
\begin{equation}
S_{\bf R}=\tau e^{i {\bf Q}\cdot{\bf R}}
\Bigl[{\case 1/2} \tilde P_0-
\sum_{m=1}^n p_m({\bf R}) \tilde P_m\Bigr]
\label{spp}
\end{equation}
and $\tilde P_m=\sum_{j=m}^n (-1)^j A_{\bf k}^{(j)}$.

The spin correlation function $S_{\bf R}$ for several $J^z/t$ values
is given in Fig.~9. We have performed the calculation for $n$ 
up to 40, which was more than sufficient to obtain converged
values. The results can be qualitatively understood visualising the
correlation function $S_{\bf R}$ in the moving coordinate frame of the
hole. For large $J^z/t$ the hole moves slowly through the N\' eel
ordered background and on the average spends more time on sublattice
$\tau$. The alternating contribution to $S_{\bf R}$ corresponds to the
AFM ordered background, and is given by the first term on the r.h.s of
Eq.~(\ref{sr}), which is apart from the AF-alternation independent of ${\bf
R}$. It represents the difference in the probability that the hole
sits on the $\uparrow$ and $\downarrow$ sublattice, respectively.
This background contribution tends to zero for $J^z/t\ll1$, where the
hole rapidly hops from one sublattice to the other.  The second term in
$S_{\bf R}$ carries all spatial dependence, i.e. defines the region of
spin disturbance, and becomes dominant at $J^z/t\ll1$.

We would like to stress here that the disappearance of the staggered
N\' eel structure for small $J^z/t$ in this correlation function is
simply a consequence of the fact that the hole visits the two
sublattices with equal probability, and it does not mean that the
antiferromagnetic order is no longer present as one could naively
conclude from similar results of a finite cluster diagonalizations.
We note that,
our results resemble surprisingly well the results for $S_{\bf R}$
obtained in exact diagonalization studies for small 
clusters\cite{bon89,hase89}. 

In Fig.~10 we show with open squares the dependence of
$S_{\bf R}$ at ${\bf R}=(1,1)$ with $J^z/t=0.4$ on the number of
magnons $n$ taken into account in the calculation. The results for
other $J^z/t$ values are in agreement with the results for ${\cal
N}_{\bf k}$, where we found that above (below) $J^z/t\sim 0.3$ a
relatively small (large) number of magnons are excited and therefore
needed in the evaluation of the correlation functions.

The conservation of the total spin $z$-component can be tested by
summing up $\sum_{R\ne 0}S_{\bf R}$. The total spin $S_{tot}^z$ is
presented in Fig.~11 as a function of $J^z/t$ with diamonds and the
full line is a guide to the eye.  $S_{tot}^z$ consists of two
parts. The first corresponds to the first term in Eq.~(\ref{spp}),
${\case 1/2} {\tilde P_0}$, and is shown with the dashed
curve. ${\tilde P_0}$ represents the difference in the probability
of the hole sitting on sublattice $\uparrow$ or 
$\downarrow$. The second term in
Eq.~(\ref{spp}) is not presented separately.  The interchange of
importance of the two contributions is in agreement with the
discussion above.  The small violation of the $S^z_{tot}$ conservation
law is a consequence of the Holstein-Primakoff representation of spin
operators.  We have also calculated $S_{tot}^z$ as a function of
$n$. For $J^z/t > 0.3$ only three magnon terms included in the wave
function give sufficient accuracy in agreement with calculation of the
norm ${\cal N}_{\bf k}$.


\subsection{Heisenberg limit ($\alpha=1$)}

The important new features of the $t$-$J$ model are (i) the
spin-dynamics described by antiferromagnetic spin waves, which have a
linear dispersion around $q=(0,0)$ and $(\pi,\pi)$, respectively. (ii)
The ground state of the model in 2D is a quantum N\'eel state, i.e.
more complex than the simple classical N\'eel ground state of the
$t$-$J^z$ model. An immediate consequence of (i) is that a
spin-deviation which is created by a single move of the hole will propagate away
from the hole in form of a spin-wave until it is reabsorbed at a later
instance.  The long wavelength spin excitations determine the
distortion of the quantum antiferromagnet at large distances from the
hole.

A further aim of our study of the RCF's is to show that the
spin correlations remain antiferromagnetic in the
vicinity of the hole. The antiferromagnetic correlations are weakened
yet not ferromagnetic. The ferromagnetic polaron picture, i.e. a
carrier accompanied by a ferromagnetically aligned spin-cluster, does
not apply here.
Ferromagnetic polarons are a quite popular scenario usually inferred by a
generalization of Nagaoka's theorem\cite{Nagaoka66}, 
which applies to the $J=0$ model,
to finite exchange interaction $J$.

To gain more insight into the complex angular dependence of the
relative correlation functions we present in addition to the numerical
results (which include up to n=3 magnons) an analytical study of the
RCF's based on the wavefunction in the one-magnon approximation. This
wave function is sufficient for a quantitative discussion in the large
$J$ case; yet it also predicts the large distance behaviour for
smaller $J$ values.

The main ${\bf k}$-dependence in the wave function stems from the
hole-magnon coupling matrix element $M_{{\bf k} {\bf q}}$.  In the $q
\to 0$ limit the ${\bf k}$- and ${\bf q}$- dependence of
$g_{{\bf k},{\bf q}} \propto q^{-1/2}(\gamma_{\bf k}-2^{1/2} {\bf
v}_{\bf k}\cdot {\bf q} / q)$ determines the asymptotic symmetry of the
correlation functions. From this structure of
$g_{{\bf k},{\bf q}}$ it is clear that at $k=0$ the
spatial symmetry is $s$-wave, whereas at the minimum of the QP band at ${\bf
k}=({\case \pi/2},{\case\pi/2})$ the symmetry is determined by the
dipolar term, where ${\bf v}_{\bf k}=\!\nabla_{\bf k}
\gamma_{\bf k}$ \cite{ram93,rei94}. 

If one is only concerned about the behaviour of the wave function at
large distance $\bf R$ from the position of the hole ${\bf R}_i$ the
one magnon contribution simplifies and one can perform the
corresponding Fourier transform of $g_{{\bf k},{\bf q}}$.
The Bloch representation of the
wave function in the limits $t/J \to 0$ and $R \to \infty$
is then approximated in leading order,
\FL
\begin{eqnarray}
|\Psi_{{\bf k} \uparrow}^{(1)}\rangle&\simeq &Z^{1/2}_{\bf k} 
\sqrt{2 \over N} 
\Biggl[\;\sum_{\,{\bf R}_i \in \;\uparrow} e^{-i{\bf k}\cdot{\bf R}_i}
h_{{\bf R}_i}^\dagger+ \nonumber\\
& &+
\!\!\sum_{{\bf R}_i \in \;\downarrow} e^{-i{\bf k}\cdot{\bf R}_i}
h_{{\bf R}_i}^\dagger
\!\sum_{\bf R} (\phi_0+i \, \phi_1) S^+_{{\bf R}_i+{\bf
R}} \Biggr]|0\rangle. \label{wf1}
\end{eqnarray}
Here the Fourier transforms
$\phi_0= -2\sqrt{2} \gamma_{\bf k} t/ (J R)$ and $\phi_1= -2({\bf
v}_{\bf k}\!\cdot\!{\bf R}) t/(J R^2)$ have different
spatial symmetries.  The $\phi_1$-term is dipolar and vanishes at
${\bf k}=(0,0)$ and $(\pi,\pi)$. At $(\frac{\pi}{2},\frac{\pi}{2})$
$\phi_1$ has its maximum, while the monopole contribution $\phi_0$
vanishes instead.  
We note that $|\Psi_{{\bf k} \uparrow}^{(1)}\rangle$ has similarity to  
the wave function describing the motion of a $^3$He atom
in superfluid $^4$He \cite{fey56,mars91}. In the following this wave
function will serve us as a
starting point for the derivation of the asymptotic properties of various
correlation functions.

The wave function, Eq.~(\ref{psi}), is properly normalized also for the
Heisenberg limit and the norm is given by Eqs.~(\ref{nak}) and
(\ref{aterms}).  The evaluation of $A_{\bf k}^{(n)}$ can be done
numerically.  In Fig.~12 $A_{\bf k}^{(n)}$ is plotted for ${\bf
k}=(k,k)$ and $n=0,1,2,3$ at $J/t=0.4$. For $n=0$, $A_{\bf
k}^{(0)}=Z_{\bf k}$ and the momentum dependence is well known\cite{mar91}. 
The next term, $n=1$, corresponds to the emission of one
magnon. The momentum dependence is very weak, which can be
qualitatively understood from the $t/J \to 0$ limit.  
For  $q < q_c \ll 1$ the one-magnon contribution
$A_{\bf k}^{(1)}$ follows as:
%
\begin{equation}
A_{\bf k}^{(1)} \Big\vert_{q_c} \simeq\,{Z_{\bf k}
\over 2 \pi^2} \!\int_0^{2 \pi}
\!\!\!\!\!d \varphi\!\!\int_0^{q_c} \!\!\!\!g_{{\bf k},{\bf q}}^2 \,q
dq \propto (\gamma_{\bf k}^2+|{\bf v}_{\bf k}|^2)\, q_c.
\label{ascaling} 
\end{equation}
Here we have put $Z_{\bf k} \sim 1$ for the weak coupling
limit. The obtained result is {\it constant} for ${\bf k}$ along
the $(1,1)$ line. This behavior is found in  the full numerical
calculations even  in the strong coupling regime $J/t=0.4$ in Fig.~12. Other
distribution functions $A_{\bf k}^{(n)}$ in Fig.~12 have a more subtle
momentum dependence which cannot be reproduced with this simple
asymptotic expansion. The sum $\sum_{n=0}^{n=3}A_{\bf k}^{(n)}$ is
close to 1, as it is clear also from Figs.~6(a,b). The results in
Fig.~12 show that the 
higher order magnon terms are less important for quasiparticle momenta 
close to the band minimum at ${\bf k}=({\case { \pi}/2},{\case {\pi}/2})$. 
For the full $J/t$ dependence of the norm at ${\bf k}=({\case
{ \pi}/2},{\case {\pi}/2})$ see Ref.~\cite{ram93}.  In order to obtain
converged results in the Heisenberg limit, we have performed
numerical calculations using unit cells with $N=16
\times 16$ up to $N=32 \times 32$. 
In summations over the Brillouin zone the points $q=0$ and ${\bf q}={\bf
Q}$ were excluded.
The numerical method of solving the SCBA equations for
$G_{\bf k}(\omega)$ was identical to Ref.~\cite{ram90}.

The average number of magnons  $\langle n \rangle$ in Fig.~13 is
presented for $J/t=0.4$ and momentum ${\bf k}=({\case {3 \pi}/8},{\case {3
\pi}/8})$, i.e. close to the QP band minimum. It is interesting that
$\langle n \rangle$ calculated for the $t$-$J$ model almost coincides
with the result obtained for the Ising case (Fig.~5).

The additional spin-deviations created by the hole motion are given by
the expression $N_{\bf R}=\langle n_0
(a^\dagger_{\bf R} a_{\bf R})\rangle-N_{\rm AFM}$.  Here we have
subtracted the large contribution $N_{AFM}=0.197$ due to the quantum
fluctuations in the ground state 
of the 2D Heisenberg antiferromagnet in the absence of the hole.  The
shape of the polaron is elongated in the direction of the QP momentum
which reflects a quasi one-dimensional motion of the polaron, as
was pointed out  in Ref.~\cite{ram93}.  
This is
consistent with the asymmetry of the QP energy band in the ``hole
pocket'' region centered around  ${\bf k}=({\case { \pi}/2},{\case {\pi}/2})$, 
where the effective next-nearest neighbor hopping for the
$(1,1)$ direction is $\sim 5
\times$ that in the $(1,-1)$ direction.
This asymmetry is most pronounced at the bottom of the QP band and
gradually vanishes away from the QP energy minimum and disappears at
$k=0$ and ${\bf k}=(\pi,0)$.  In the limit $R \to \infty$ the
perturbative result is to lowest order in $t/J$ given by 
\begin{equation}
N_{\bf
R}= {8 t^2 \over J^2 R^2}
(\gamma_{\bf k}^2+({{\bf v}_{\bf k}\!\cdot\!{\bf R} \over R})^2).
\end{equation}
This result strictly holds only asymptotically, but nevertheless
it reflects all symmetries found at short distances in the
numerical treatment. The momentum dependence is qualitatively correct
as well,
while the $J/t$ dependence is correct only in the range of
validity of perturbation, $t/J \to 0$. The correlation function
$N_{\bf R}$ decays as a power-law, $N_{\bf R} \propto R^{-2}$.
Although the number of excited magnons $\langle n \rangle$ is small,
it turns out that the change in the total number of spin deviations
$\sum_{\bf R}N_{\bf R}$ diverges logarithmically. The definition of
the polaron size used for the Ising limit of the model thus cannot be
used here. 
Since the magnetic excitations $\omega_{\bf q}$ vanish
linearly with $q$, also other correlation functions show power
law decay, yet with different exponents\cite{ram93}.

In Fig.~14 we display the distribution of $z$ component of spin $S_{\bf
R}$ as a function of $J/t$ and for ${\bf k}=({\case {3 \pi}/8},{\case
{3 \pi}/8})$. 
This correlation function depends strongly on the direction and size 
of the momentum of the quasiparticle  (see Ref.~\cite{ram93}).
The asymmetry of the
polaron is reflected  in different values for $S_{\bf R}$ at positions
${\bf R}$ labelled with 2 and 2' or 4 and 4', respectively. 
This result is quite different from the isotropic perturbation in
the Ising limit (e.g., Fig.~9).  We stress that the same
asymmetry was found in numerical studies of an 18 site $t$-$J$ cluster
with one hole\cite{els90}.  The ground state is at ${\bf k}=({\case
2/3} \pi,{\case 2/3} \pi)$ for $J/t=0.4$ [The point ${\bf k}=({\case
\pi/2},{\case \pi/2})$ is absent in that system].  Due to the high
symmetry of the $4\times 4$ cluster such subtle asymmetries of the
polaron cannot be studied there. The $J/t$ dependence of $S_{\bf R}$
is in excellent agreement with the results of
Refs.~\cite{bon89,hase89}. As finite size effects in such small
clusters are expected to be quite large, the good agreement of $S_{\bf R}$
with the exact results is surprising. The reason for the
disappearance of the AFM structure in this correlation function
for small $J/t$ is as in the Ising limit a
consequence of fast hole motion which leads to an average over the two
sublattices. It
does not imply that the antiferromagnetic order of the spin background
is destroyed.  The correlation function is small in this limit because the
polaron is large and many  sites contribute to
the sum rule $S^z_{tot}= {\case 1/2}$.

To test our analytical and  numerical procedure we have calculated 
$S_{\bf R}$ for different number of magnons terms $n$ (Fig.~10).
The convergence rate is similar as in the case of Ising limit,
i.e. three magnon terms give a sufficient accuracy. 
To display the anisotropy  $S_{\bf R}$ is shown in Fig.~10 for
${\bf R}=(1,\pm 1)$ and ${\bf k}=({\case {\pi}/2},{\case {\pi}/2})$
with $J/t=0.4$ as a function of number of magnon
lines in the wave function, $n=1,2,3$. The corresponding contributions
(diagrams) are labelled with symbols $(B_n)$ and ($C_{nm}$) as defined in 
Fig.~7. The asymmetry of the polaron, which is fully consistent with the 
numerical results
of Ref.~\cite{els90}, can be attributed to the diagram $C_{02}$, corresponding
to a two magnon process.

The effect of the hole on the AF correlations and the energy of the
spin system is measured by the nearest-neighbor spin correlation
function $C_{\bf R}=
\langle n_0 \,({\bf S}_{{\bf R}_1}\!\cdot\!{\bf S}_{{\bf R}_2})\rangle$ 
defined on bonds between two neighboring sites $(1-2)$, ${\bf R}=({\bf
R}_1+{\bf R}_2)/2$ \cite{bon89}. In Fig.~15(a) $C_{\bf R}$ is shown as
a function of $J/t$ and for ${\bf k}=({\case {3 \pi}/8},{\case {3
\pi}/8})$.  The correlation function remains negative and in agreement
with the numerical result obtained on a 16 sites cluster\cite{bon89}.
Hence AFM-correlations persist in the vicinity of the hole contrary to
what one would expect from the ferromagnetic polaron picture.  
$C_{\bf R}$ is asymmetric as can be seen,
e.g., from the bonds ${\bf R}=(1,\pm {\case 1/2})$, labeled with 1 and
$1'$. The momentum dependence of $C_{\bf R}$ can be explained
with the perturbative result which follows from
asymptotic wave function Eq.~(\ref{wf1}) \cite{ram93}
%
\begin{equation}
C_{\bf R}\!= -0.329+
{4 t^2 \over J^2 R^4}
(\gamma_{\bf k}^2+2 |{\bf v}_{\bf k}|^2).
\end{equation}
This correlation function decays as $R^{-4}$ at large distances and
could be used as a definition of the size of the polaron.  Our results
suggest that the size of the polaron measured with this correlation
function is, at moderate $J/t=0.4$, of the order of a few lattice
sites.

Another interesting aspect of the deformation of the spin background
is contained in the bond-spin currents ${\bf j}_{\bf R}=
\langle n_0 ({\bf S}_{{\bf R}_1} \!\times\! {\bf S}_{{\bf R}_2})^z 
{\bf u} \rangle$, where ${\bf u}$ is a unit vector ${\bf
u}={\bf R}_2-{\bf R}_1$ \cite{els90,inu90}.  This quantity
follows from the equation of motion for the spin density
%
\[
\dot{\bf S}_{\bf R}
\!=\!it \! \sum_{{\bf u},s s'}\!(\hat{\sigma}_{s s'}
c^{\dagger}_{{\bf R},s}  c_{{\bf R+u},s'} \!-
\!{\rm H.c.}) \!-\! 
2iJ \!\sum_{\bf u}
{\bf S}_{\bf R} \!\times\! {\bf S}_{\bf R \!+\!u},
\]
%
where $\hat{\sigma}$ are Pauli spin matrices. Here the first term is
the spin current induced by the hopping of the hole and the second
term ($\sim {{\bf j}_{R}}$) describes the backflow in the spin
system. Due to the broken symmetry total spin is not a good quantum
number, therefore we consider only the $z$ component of the current.
In Fig.~15(b) ${\bf j}_{\bf R}$ is presented as a function of $J/t$
and for ${\bf k}=({\case {3 \pi}/8},{\case {3 \pi}/8})$.  ${\bf
j}_{\bf R}$ is an odd function with respect to the wave vector (at
$k=0$). Because of symmetry it vanishes also at ${\bf
k}=(\pi,0)$. Since the ground state has AFM long-range order, the
points ${\bf k}$ and ${\bf k}+(\pi,\pi)$ are equivalent, and
therefore, ${\bf j}_{\bf R}$ vanishes also at ${\bf k}\!=\!({\case
\pi/2},{\case \pi/2})$.  Comparison of ${\bf j}_{\bf R}$ with exact
diagonalization results is delicate. As reported in 
Ref.~\cite{ram93} we find good agreement with results from
Ref.~\cite{els90}.  For the complete momentum dependence of ${\bf j}_{\bf R}$
in comparison with exact diagonalization see
Ref.~\cite{jrh94}. Agreement is excellent in the anisotropic limit,
$\alpha < 1$. In the Heisenberg limit a reliable comparison is very
difficult because of the strong ${\bf k}$ dependence of ${\bf j}_{\bf R}$
which makes it very sensitive to the boundary conditions of small
clusters. In Fig.~15(b) we present ${j}_{\bf R}$ for various bonds
${\bf R}$ defined in Fig.~15(a). 
The asymptotic pattern of bond spin currents is dipolar
\cite{ram93},
%
\begin{equation}
{j}_{\bf R}={4\sqrt{2}
t^2 \over J^2 R^3}
\gamma_{\bf k}\,[{\bf v}_{\bf k}-{({\bf v}_{\bf k}\!\cdot\!{\bf  R})\, 
{\bf R} \over R^2}]\!\cdot\!{\bf u}. 
\end{equation}
The spin backflow
current ${\bf j}_{\bf R}$ decays as $R^{-3}$ and vanishes in the
ground state for ${\bf k}=({\case \pi/2},{\case \pi/2})$.
Again the general momentum dependence is correct as in the case of the
other correlation functions considered.

\section{CONCLUSIONS}

We have outlined a method that allows to calculate the real-space
structure of a spin polaron in a quantum antiferromagnet. The approach
is based on the spin-polaron formulation for the $t$-$J$ model, where
holes are described as spinless fermions while the spin excitations
are treated within linear spin-wave theory. The single-particle
Green's function in (selfconsistent) Born approximation, which has
been shown earlier to provide an excellent description of the
numerical data for the $t$-$J$ model, is sufficient to calculate the
many-body wave function describing the polaron (Reiter's wave
function). We have shown here how this wave function can be used 
to calculate in the frame of the selfconsistent Born approximation
quite complex correlation functions. 

Our calculation of a number of correlation functions, which measure
the deformation of the spin system due to a moving hole (spin
polaron), provides a very detailed check of this approach against
exact diagonalization. In particular we have shown that the 
spectral weight of the spin-polaron quasiparticle calculated from the
wave function is consistent with the result derived from the Green's
function. It is demonstrated how the number of spin-excitations
involved in the polaron formation increases with decreasing spin
stiffness when $J\rightarrow 0$.  
 We have
determined the probability distribution of the number of magnons
excited in the ground state and found that for
$J/t=0.4$ (a typical value for copper-oxide superconductors)
the average number of magnons is about one. In  the
Ising limit the average number of magnons scales as
$\langle n \rangle \propto (t/J^z)^{1/3}$ 
as $J^z/t \to 0$ in agreement with the string
picture for the moving hole and with the Ref.~\cite{mat91}.

We have put particular emphasis on the asymptotic decay of the
perturbations introduced by the spin polaron formation. 
Since the spin wave energies $\omega_{\bf q}$ in a quantum antiferromagnet
vanish linearly with $q$, perturbations  
in the spin system decay with a power law.
For example the change of the local spin deviations $N_{\bf R}\propto
R^{-2}$, while the perturbation of the nearest-neighbor spin
correlations decays as $R^{-4}$ with the distance from the hole. 
In the $t$-$J$ model all correlation functions have a quite complex
structure in real space which depends on the momentum of the polaron,
whereas in the $t$-$J^z$ model all perturbations are isotropic.

We note that despite of the power law decay of the polaron correlation
functions in the $t$-$J$ model the quasiparticle spectral weight does
not vanish. Whether this is correct or
an artifact of the selfconsistent Born approximation remains to be
shown by a more rigorous treatment.

Finally we want to stress that the approach discussed here may also be
applied to other interesting problems such as strongly correlated
electrons coupled to Holstein or other phonons.

We acknowledge useful discussions with G. Khaliullin, A. M. Ole\' s, P.
Prelov\v sek and I. Sega
and thank G. Mart\'{\i}nez for
the careful reading of the manuscript. One of us (A.R.) would like to 
thank L. Hedin and
the Max-Planck-Institut FKF for the hospitality extended to him
during several stays. We also acknowledge
financial support from BMBF (Bonn) under project-no. SLO-007-95. 



\begin{figure}
\caption{Diagrammatic representation of the wave function 
$|\Psi_{{\bf k}\tau}^{(n)}\rangle$ for $\tau=\uparrow$. (a) The first
three terms contain no-magnon, one-magnon and two-magnons excitations,
respectively.
(b) The double line represents the single particle Green's function
in noncrossing approximation.  \label{fig1}}
\end{figure}

\begin{figure}
\caption{Diagrammatic representation of the norm ${\cal N}_{\bf k}$. The
first term -- containing no magnon line -- is identical to the QP pole
strength $Z_{\bf k}$. \label{fig2}}
\end{figure}

\begin{figure}
\caption{The norm ${\cal N}_{\bf k}$ in the Ising limit as a function
of the number, $n$, of magnons included in the wave function for various 
$J^z/t$. \label{fig3}}
\end{figure}

\begin{figure}
\caption{The distribution of the number of magnons  $A_{\bf
k}^{(n)}$ as a function of $n$ for various $J^z/t$. \label{fig4}}
\end{figure}

\begin{figure}
\caption{Average number of magnons $\langle n \rangle$ in the Ising
limit as a function of $J^z/t$. The inset shows $\langle n \rangle$ on
logarithmic scale compared with asymptotic results. \label{fig5}}
\end{figure}

\begin{figure}
\caption{The norm ${\cal N}_{\bf k}$ for the $t$-$J$ model as a function
of $J/t$ including magnons up to order $n=3$ in the quasiparticle
wavefunction. (a) Momentum at the top of the QP band at ${\bf k}=0$, and
(b) at ${\bf k}=({\case {3 \pi}/8},{\case {3 \pi}/8})$ , i.e.
close to the QP-band minimum. The numerical calculation was performed for a
momentum space
grid corresponding to a $32 \times 32$ system. The solid and the dashed lines
are guides to the eye only. \label{fig6}}
\end{figure}

\begin{figure}
\caption{Diagrammatic representation of correlation functions. Each
class of $(B_n)$ diagrams contains $n$ noncrossing diagrams. 
$(C_{nm})$ diagrams appear always in pairs with
the H.c. counterparts. For a detailed description see text.
\label{fig7}}
\end{figure}

\begin{figure}
\caption{Radius of the polaron in the Ising limit vs. $J^z/t$ for
various definitions: $R_p$ for $p=0.75$ (full circles),
$p=0.90$ (open squares), and full squares for $p=0.99$. 
Here $p$ defines the fraction of spin deviations within the radius
$R_p$. The full line
represents root mean square radius, $R_{RMS}$, while the dashed line
gives the average radius $\langle R \rangle$.\label{fig8}}
\end{figure}

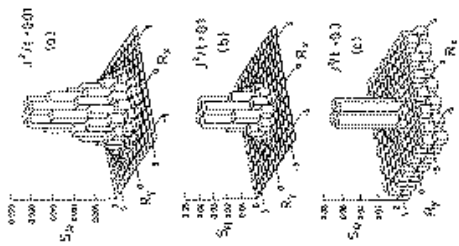
\begin{figure}
\caption{Distribution of the $z$-component of spin 
$S_{\bf R}=\langle n_0 S_{\bf R}^z \rangle$ around
the moving hole for three different values for $J^z/t$.\label{fig9}}
\end{figure}

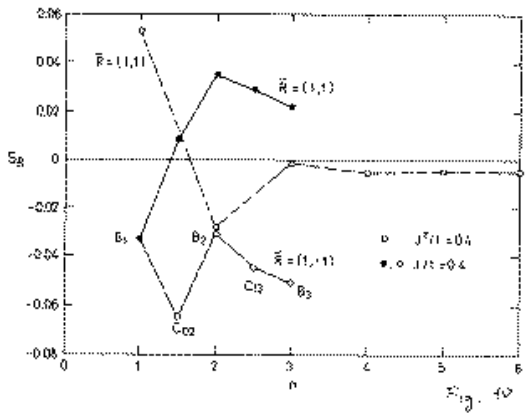
\begin{figure}
\caption{The dependence of $S_{\bf R}$ on the number of magnon terms $n$ 
in the wave function. In the Ising limit $S_{(1,1)}$ (dash-dotted line)
is essentially converged for $n>3$ given $J^z/t=0.4$. In the
Heisenberg case the contributions from different diagrams to
$S_{\bf R}$ are shown for ${\bf R}=(1,1)$ (solid) and
${\bf R}=(1,-1)$ (dashed line), respectively, for $J/t=0.4$
and ${\bf k}=({\case {\pi}/2},{\case {\pi}/2})$. 
For the classification of diagrams $(B_n)$ and $(C_{nm})$ 
see Fig.~7.\label{fig10}}
\end{figure}

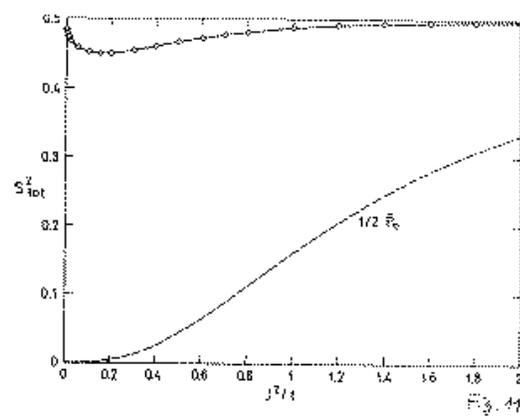
\begin{figure}
\caption{The total $z$-component of spin $S_{tot}^z$ vs. $J^z/t$
(diamonds). The dashed line represents ${\case 1/2} \tilde P_0$
in Eq.~(36),
where $\tilde P_0$ is the difference of the probabilities
for  the hole to occupy sublattice $\uparrow$ or 
$\downarrow$, respectively.\label{fig11}}
\end{figure}

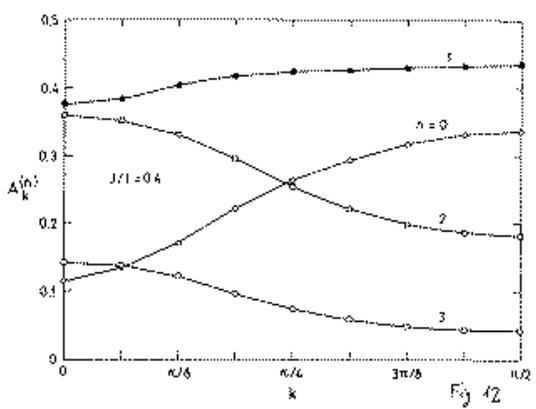
\begin{figure}
\caption{Magnon distribution function $A_{\bf k}^{(n)}$ for the
$t$-$J$ model ($J/t=0.4$) 
as function of ${\bf k}=(k,k)$
for various magnon numbers
$n$: $n=0$ (diamonds) $n=1$, (full circles), $n=2$ (open
squares), and $n=3$ (open circles). The numerical calculation of all
matrix elements was performed on a grid corresponding to a $16 \times
16$ system.  Lines connecting the symbols are guides to the eye only.
\label{fig12}}
\end{figure}

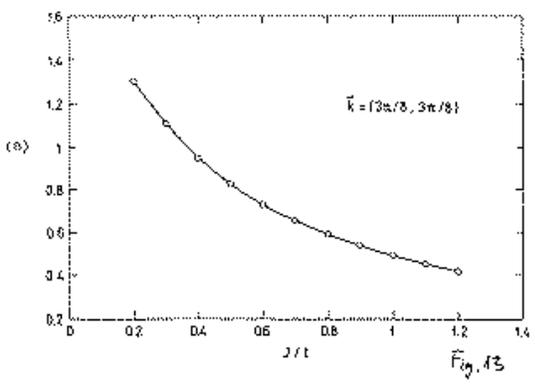
\begin{figure}
\caption{The average number of magnons
$\langle n \rangle$ involved in the spin polaron formation 
in the $t$-$J$ model as function of $J/t$.
The polaron momentum is ${\bf
k}=({\case {3 \pi}/8},{\case {3 \pi}/8})$.
\label{fig13}}
\end{figure}

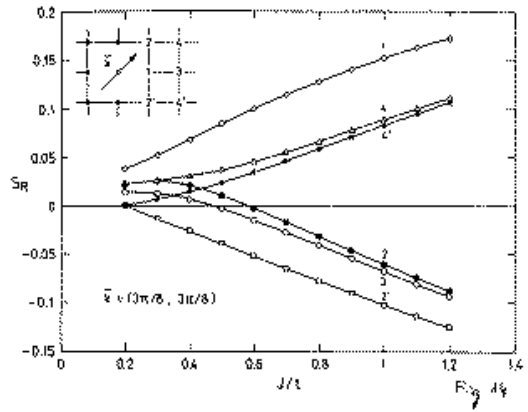
\begin{figure}
\caption{The $z$-component of the spin correlation function 
$S_{\bf R}$ vs. $J/t$ for
${\bf k}=({\case {3 \pi}/8},{\case {3 \pi}/8})$. Note the asymmetry
between the directions ${\bf R}\| (1,1)$ and ${\bf R}\|
(1,-1)$ in the $t$-$J$ model.  The numerical calculation was
performed on a grid corresponding to $16 \times 16$ system. \label{fig14}}
\end{figure}

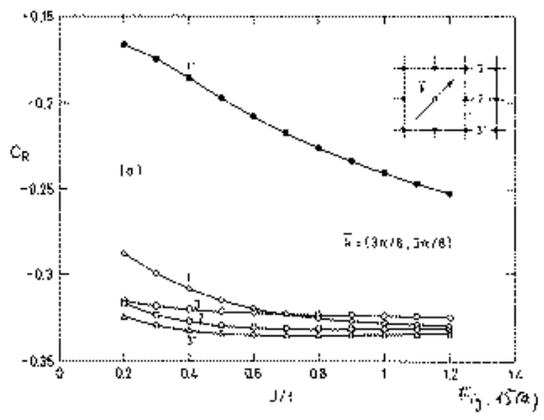
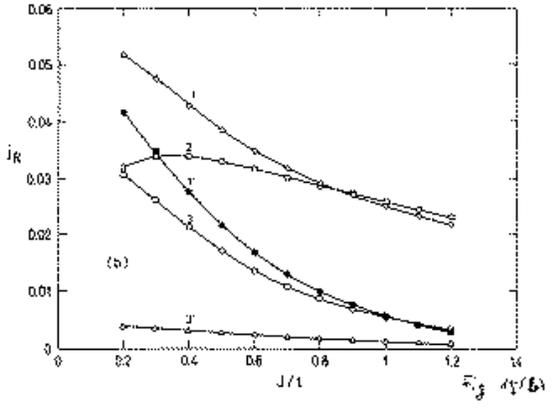
\begin{figure}
\caption{(a) Nearest-neighbor spin-correlation
function $C_{\bf R}$ and  (b) the $z$-component of the
bond spin currents ${\bf j}_{\bf R}$
as function of $J/t$ for the quasiparticle momentum ${\bf
k}=({\case {3 \pi}/8},{\case {3 \pi}/8})$.  
The inset in (a) provides a definition of the $n.n.$-correlations considered.
In both (a) and (b) note the
asymmetry between the directions ${\bf R}\| (1,1)$ and ${\bf
R}\| (1,-1)$. The asymptotic behavior is given by Eqs. (40) and (41),
respectively. 
\label{fig15}}
\end{figure}


\begin{references}    

\bibitem{rev} {For recent reviews see: 
E. Dagotto, Rev. Mod. Phys. {\bf 66}, 763 (1994); 
Yu. A. Izyumov, Physics - Uspekhi {\bf 40} (5), 445, (1997).}

\bibitem{hor89} 
{K. J. von Szczepanski, P. Horsch, W. Stephan, and M. Ziegler,
Phys. Rev. B {\bf 41}, 2017 (1990);
P. Horsch {\it et al.}, Physica C {\bf 162-164}, 783 (1989).}

\bibitem{bon89} {J. Bon\v ca, P. Prelov\v sek, and I. 
Sega, Phys. Rev. B {\bf 39}, 7074 (1989).}

\bibitem{hase89} {Y. Hasegawa and D. Poilblanc, Phys. Rev. B
{\bf 40}, 9035 (1989).}

\bibitem{Dagotto90}
{E. Dagotto, R. Joint, A. Moreo, S. Bacci, and  E. Gagliano,
 Phys. Rev. B {\bf 41}, 9049 (1990).}

\bibitem{poi92} {D. Poilblanc, T. Ziman, 
H. J. Schulz, and E. Dagotto, Phys. Rev. B {\bf 47}, 14267 (1993).}

\bibitem{Leung95}
{P. W. Leung and R. J. Gooding, Phys. Rev. B {\bf 52}, R15711 (1995).} 

\bibitem{sch88} {S. Schmitt-Rink, C. M. Varma, and A. E. Ruckenstein,
Phys. Rev. Lett. {\bf 60}, 2793  (1988).}

\bibitem{kan89} {
C. L. Kane, P. A. Lee, and N. Read, Phys. Rev. B {\bf 39}, 6880 (1989).} 

\bibitem{mar91} {G. Mart\'{\i}nez and P. Horsch, Phys. Rev.  B {\bf 44},
317 (1991). }

\bibitem{Igarashi92}
{J. Igarashi and P. Fulde, Phys. Rev. B {\bf 45}, 10419 (1992).} 

\bibitem{tru88} {S. A. Trugman, Phys. Rev. B {\bf 37}, 1597 (1988).}

\bibitem{ede90} {R. Eder and K. W. Becker, Z. Phys. B {\bf 78}, 219 (1990).}

\bibitem{Prelovsek90}   
{P. Prelov\v sek, I. Sega, and J. Bon\v ca, 
Phys. Rev. B {\bf 44}, 317 (1990).} 

\bibitem{su89} {Z. B. Su, Y. M. Li, W. Y. Lai, and L. Yu, Phys. Rev.
Lett. {\bf 63}, 1318 (1989).}

\bibitem{Riera97}
{J. A. Riera and E. Dagotto, Phys. Rev. B {\bf 55}, 14543 (1997).}

\bibitem{Barentzen96}
{H. Barentzen, Phys. Rev. B {\bf 53}, 5598 (1996).}

\bibitem{els90} {V. Elser, D. A. Huse, B. I. Shraiman, and E. D.
Siggia, Phys. Rev. B {\bf 41}, 6715 (1990).}

\bibitem{shr88} {B. I. Shraiman and E. D. Siggia, Phys. Rev. Lett. {\bf
60}, 740 (1988); {\bf 61}, 467 (1988).}

\bibitem{ram93}
{A. Ram\v sak and P. Horsch, Phys. Rev. B {\bf 48}, 10559 (1993).}

\bibitem{Wells95}
{B. O. Wells {\it et al.}, Phys. Rev. Lett. {\bf 74}, 964 (1995).}

\bibitem{tprime}
{T. Tohyama and S. Maekawa, Phys. Rev. B {\bf 49}, 3596 (1994);
J. Ba\l a and A. M. Ole\' s, Phys. Rev. B {\bf 52}, 4597 (1995);
A. Nazarenko {\it et al.}, Phys. Rev. B {\bf 51}, 8676 (1995);
B. Normand and P. A. Lee, Phys. Rev. B {\bf 51}, 15519 (1995);
V. I. Belinicher {\it et al.} Phys. Rev. B {\bf 54}, 14914 (1996);
N. M. Plakida {\it et al.} Phys. Rev. B {\bf 55}, R11997 (1997).
}
\bibitem{complexH} 
{O. A. Starykh {\it et al.} Phys. Rev. B {\bf 52}, 12534 (1995);
J. Ba\l a, A. M. Ole\' s and J. Zaanen, Phys. Rev. B {\bf 54}, 10161
(1996); F. Lema and A. A. Aligia,  Phys. Rev. B {\bf 55}, 14092
(1997).}

\bibitem{bul68} {L. N. Bulaevski, E. L. Nagaev, and D. I. Khomskii,
Zh. Eksp. Teor. Fiz. {\bf 54}, 1562 (1968) [Sov. Phys. JETP {\bf 27},
836 (1968)].}

\bibitem{bri70} {F. Brinkman and T. M. Rice, Phys. Rev. B {\bf 2},
1324 (1970).}

\bibitem{and87} {P. W. Anderson, Science {\bf 235}, 1196 (1987).}

\bibitem{and90} {P. W. Anderson, Phys. Rev. Lett. {\bf 64}, 1839 (1990).}

\bibitem{weng97} {Z. Y. Weng, D. N. Sheng, Y. C. Chen, and C. S. Ting,
Phys. Rev. B {\bf 55}, 3894 (1997).}

\bibitem{Plakida94}
{N. M. Plakida, V. S. Oudovenko, and V. Yu. Yushankhai, 
  Phys. Rev. B {\bf 50}, 6431 (1994).}

\bibitem{Sherman94}
{A. Sherman and M. Schreiber,  Phys. Rev. B {\bf 50}, 12887 (1994).}

\bibitem{Plakida97}
{N. M. Plakida, V. S. Oudovenko, P. Horsch, and A. I. Liechtenstein, 
  Phys. Rev. B {\bf 55}, R11997 (1997).}

\bibitem{rei94} {G. F. Reiter, Phys. Rev. B {\bf 49}, 1536 (1994).} 

\bibitem{ram92} {A. Ram\v sak, P. Horsch, and P.
Fulde, Phys. Rev. B {\bf 46}, 14305 (1992).}

\bibitem{kyu96} {B. Kyung, S. I. Mukhin, V. N. Kostur and R. A. Ferrell, Phys. 
Rev. B {\bf 54}, 13167 (1996).}

\bibitem{Horsch94}
P. Horsch and A. Ram\v sak, J. Low Temp. Phys. {\bf 95}, 343 (1994).

\bibitem{ram90} {A. Ram\v sak and P. Prelov\v sek, Phys. Rev. B
{\bf 42}, 10415 (1990).} 

\bibitem{bul89} {N. Bulut, D. Hone, D. J. Scalapino, and E. Y. Loh,
Phys. Rev. Lett. {\bf 62}, 2191 (1989).}


\bibitem{mal92} {A. G. Mal'shukov and G. D. Mahan, Phys. Rev.
Lett. {\bf 68}, 2200 (1992).}

\bibitem{liu92} {Z. Liu and E. Manousakis, Phys. Rev. {\bf B} 45,
2425 (1992).}

\bibitem{dag91} {E. Dagotto and J. R. Schrieffer, Phys. Rev. B {\bf
43}, 8705 (1991).}

\bibitem{mat91} {D. C. Mattis and H. Chen, 
Int. J. Mod. Phys. B {\bf 5}, 1401 (1991).}

\bibitem{Nagaoka66}
{Y. Nagaoka, Phys. Rev. {\bf 147}, 392 (1966).}

\bibitem{fey56} {R. P. Feynman and M. Cohen, Phys. Rev. {\bf 102},
1189 (1956); W. L. McMillan, {\it ibid.} {\bf 175}, 266 (1968).}

\bibitem{mars91} {F. Marsiglio, A. E. Ruckenstein, S. Schmitt-Rink, and
C. M. Varma, Phys. Rev. B {\bf 43}, 10882 (1991).}

\bibitem{inu90} {J. Inoue and S. Maekawa, J. Phys. Soc. Jpn. {\bf 59},
2119 (1990); {\bf 59}, 3467 (1990).}

\bibitem{jrh94} {A. Ram\v sak, P. Horsch and J. Jakli\v c,
Physica C {\bf 235-240}, 2247 (1994).}



\end{references}
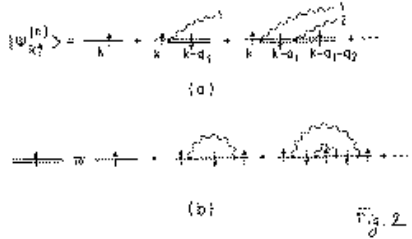
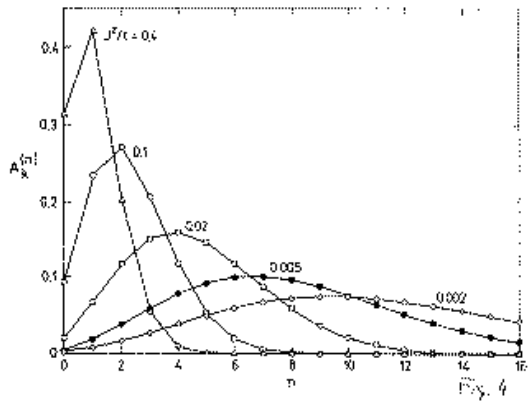
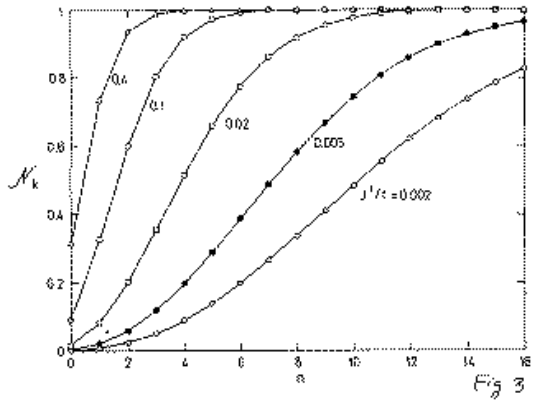
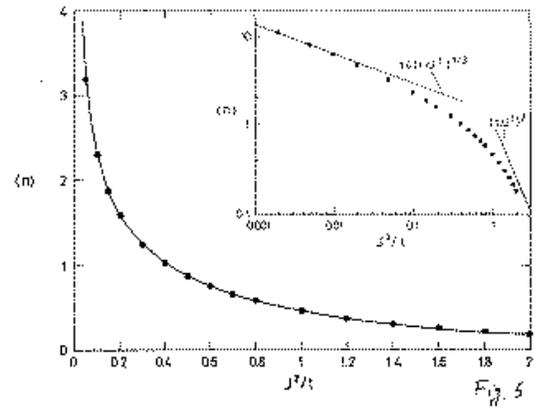
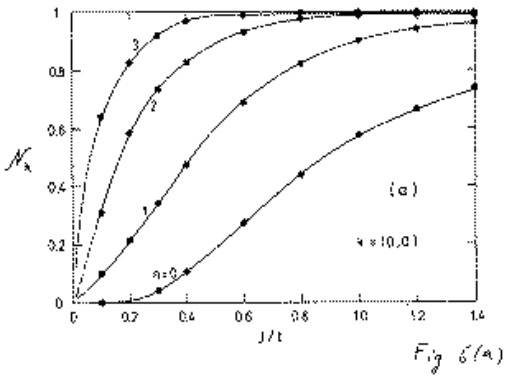
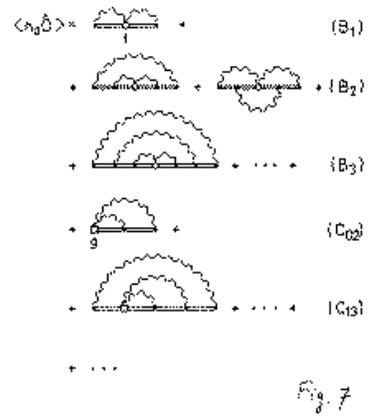
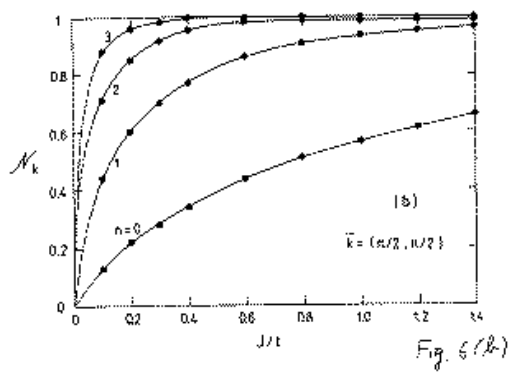
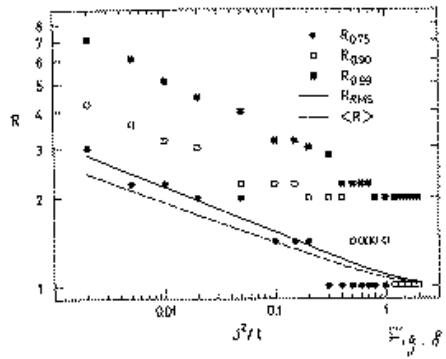
\end{document}